%% file: paper.tex
\documentclass[10pt,conference]{IEEEtran}
\IEEEoverridecommandlockouts
\usepackage{cite}
\usepackage{amsmath,amssymb,amsfonts}
\usepackage{algorithmic}
\usepackage{graphicx}
\usepackage{textcomp}
\usepackage{xcolor}
\usepackage{diagbox}
\usepackage{multirow}

\newcommand{\quotes}[1]{``#1''}

\usepackage{array}

\usepackage{enumitem}
\newlist{Oenumerate}{enumerate}{1}
\setlist[Oenumerate]{label=O.\arabic*}
\usepackage[draft,inline,nomargin,index]{fixme}  \fxsetup{theme=color,mode=multiuser} \FXRegisterAuthor{ft}{envft}{\color{red}FT} \FXRegisterAuthor{fm}{envfm}{\color{blue}FM} \FXRegisterAuthor{lo}{envlo}{\color{orange}LO} \fxsetup{layout={footnote}} \fxusetheme{colorsig}

\makeatletter
\def\@IEEEtablestring{figure}
\makeatother

\def\BibTeX{{\rm B\kern-.05em{\sc i\kern-.025em b}\kern-.08em
    T\kern-.1667em\lower.7ex\hbox{E}\kern-.125emX}}
\begin{document}

\author{
    \IEEEauthorblockN{Fadi Mohsen, Loran Oosterhaven and
    Fatih Turkmen}
    \IEEEauthorblockA{Bernoulli Institute for Mathematics, Computer Science and Artificial Intelligence\\ University of Groningen, The Netherlands\\
    f.f.m.mohsen@rug.nl, l.oosterhaven@student.rug.nl, f.turkmen@rug.nl}
}

\title{KotlinDetector: Towards Understanding the Implications of Using Kotlin in Android Applications
}

\maketitle

\begin{abstract}
Java programming language has been long used to develop native Android mobile applications. 
In the last few years many companies and freelancers have switched into using Kotlin partially or entirely. 
As such, many projects are released as binaries and employ a mix of Java and Kotlin language constructs. 
Yet, the true security and privacy implications of this shift have not been thoroughly studied. 
In this work, a state-of-the-art tool, \emph{KotlinDetector}, is developed to directly extract any Kotlin presence, percentages, and numerous language features from Android Application Packages (APKs) by performing heuristic pattern scanning and invocation tracing. Our evaluation study shows that the tool is considerably efficient and accurate. We further provide a use case in which the output of the \emph{KotlinDetector} is combined with the output of an existing vulnerability scanner tool called \emph{AndroBugs} to infer any security and/or privacy implications. 

\end{abstract}


\input{Introduction.tex}
\input{Background.tex}
\input{Related.tex}
\input{Approach.tex}

\input{Implementation.tex}
\input{Evaluation.tex}

\input{UseCase.tex}
\input{Conclusion.tex}

\input{ack.tex}

\bibliographystyle{IEEEtran}
\bibliography{paper}

\end{document}

%% file: Introduction.tex
\section{Introduction}
One relatively new development in the world of mobile Android applications is the \emph{Kotlin} programming language. In 2017, Google officially started supporting Kotlin in the Android Studio IDE \cite{kotlinsupport}. Even more recently, in 2019, Google announced that Kotlin now is the preferred programming language for Android applications \cite{kotlinpreferred}.
Kotlin is a cross-platform programming language developed and introduced by JetBrains. It is designed to seamlessly work together with Java and thus, like Java, heavily relies on the Java Virtual Machine. Originally, Java was the preferred programming language for Android applications. It is due to the dependency of Android on Java and the interoperability between Java and Kotlin, that Kotlin became the most promising replacement for the preferred language of Android applications. The main features of the Kotlin programming language include null safety, expressive code, anonymous and high-order functions and interoperability. Overall, Google claims that using Kotlin for mobile applications results in more modern, expressive and safer code \cite{kotlinhome}.

To the best of our knowledge, there have been no thorough and large-scale studies to investigate the implications of shifting towards Kotlin for mobile application development.
This is mainly due to the fact that Kotlin is a relatively new language and there is a lack of tools, tagged datasets and open source Android projects. 
For example, F-droid contains a repository of free and open source Android apps \cite{fdroidstats}. The F-Droid repository contains a growing number of nearly 3,000 apps \cite{F-DroidRepo}, compared to 2.87 million apps on the Google Play Store \cite{Playstorerepo}. The source codes of some of these apps are hosted on GitHub, which shows the languages used in these codes via their Linguist tool \cite{linguist}. 

The GitHub Linguist tool on the other hand, which was first introduced by GitHub in 2011, is used to identify the programming language in code repositories published on their platform including Android projects. Mateus and Martinez \cite{studyqualitykotlin} performed an empirical study on the quality of Android applications written in the Kotlin programming language. Their approach relies on first determining whether an application uses Kotlin or not and if it does then the percentage of Kotlin code that is present in the application is estimated. Both of these approaches assume full access to the original source code, which is not the case with the vast majority of Android mobile applications. In addition, the Kotlin features are not retrieved, but only the percentages. The percentages do not take into the account that some source code might be skipped by the compiler. 

Thus, the aim of this work is threefold: first, build a tool to detect the presence of Kotlin in Android Application Packages (APKs), estimate Kotlin percentage in these APKs, and extract numerous language features. Second, provide a usage scenario for our tool, in which its output is combined with the output of existing vulnerability scanning tool called \emph{AndroBugs} \cite{AndroBugs15} to infer the security and privacy implications of using Kotlin in different datasets. Third, prepare, validate and disseminate tagged Kotlin related datasets.

Our tool, \emph{KotlinDetector}, requires no access to the source code (i.e. black-box), instead it extracts and scans the \emph{dex} and \emph{AndroidManifest.xml} files even if an obfuscation is used. It employs a number of techniques such as invocation tracing, heuristic pattern scanning and heuristic searching. In evaluating the efficiency and accuracy of \emph{KotlinDetector}, we curated several datasets of Kotlin related Android applications that we collected, built or borrowed from the literature. 
Our evaluations show that \emph{KotlinDetector} is both efficient and accurate, which makes it a reliable source for conducting various Kotlin related studies such as the privacy and security implications of Kotlin usage in mobile application development.

The remainder of the paper is organized as follows. We first start by providing context for the information discussed throughout the paper in Section~\ref{sec:bg} and discuss the related works in Section~\ref{sec:related_work}. The approach
and the implementation details are explained in Section~\ref{sec:approach} and \ref{sec:implementation} respectively. Section~\ref{sec:evaluation} presents the details of our evaluation of \emph{KotlinDetector} while Section~\ref{sec:usagescenario} provides an example study during which \emph{KotlinDetector} can be used. Finally, we conclude the paper in Section~\ref{sec:conclusion}.

%% file: Background.tex
\section{Background}
\label{sec:bg}
Android applications are published and shared world-wide in the Android Package file format (\texttt{.apk}). It is quite similar to a JAR file in the sense that they are both compressed archives using the ZIP format. Nevertheless, their contents are different. Any APK file usually contains the following files: \texttt{AndroidManifest.xml},
\texttt{classes.dex}, \texttt{resources.arsc}, \texttt{res}, \texttt{assets}, \texttt{lib}, and \texttt{META-INF}. Before we provide the implementation details of the \emph{KotlinDetector} tool, we first start with explaining multi-dex files, obfuscation, and compiled Kotlin.

\subsection{Multi-DEX}
\label{sec:multidex}
Most small applications consist of just one DEX file. Modern applications tend to be larger, may consist of multiple DEX files.  
Method indexes or string indexes consisted of just two bytes. This imposes a limitation on how many methods can be referenced inside a single DEX file, namely 64K. Frequently, large applications consist of more than 64K methods and therefore they cannot all be possibly referenced using the current Dalvik instruction set. 
To overcome this issue multi-DEX applications were introduced. In multi-DEX applications, there are other DEX files besides the usual \texttt{classes.dex} file. An arbitrary amount of additional DEX files can exist within a multi-DEX application. They are named \texttt{classesN.dex}, starting with \texttt{N=2}. Each DEX file can reference up to 64K methods on its own. The method references are balanced out over all DEX files to ensure this limit is not reached. Upon installation, the Android Runtime compiles all the DEX files combined.

Multi-DEX applications is also the reason that a method that is referenced inside a DEX file, does not have to be defined in this DEX file, but can actually be defined in another DEX file. Multi-DEX applications brought many challenges to the design and implementation of \emph{KotlinDetector}. These will be discussed in Section~\ref{sec:implementation}.

\subsection{Obfuscation}
\label{sec:background_obfuscation}

Obfuscation, in the context of software, is the act of deliberately transforming source code or machine code into a form which is hard for humans to understand. Hence, the development of code obfuscation techniques is frequently driven by the desire to hide specific implementation details from automated or human analysis~\cite{obfdef}. This is particularly important for software developed for the Android Runtime platform, due to the fact that type information and variable name information are present in the applications' DEX files. Obfuscation has always been popular among malware developers, but is also used by legitimate companies to protect their intellectual property. 

In the context of Android there are four popular obfuscation techniques, namely identifier renaming, string encryption, Java reflection and packing. Following is a brief discussion of the identifier renaming, which is the mostly used obfuscation technique among Android developers \cite{androidobf}, and the one that is currently supported by \emph{KotlinDetector}.

\subsubsection{Identifier renaming}
The names and types of variables, fields and methods of an Android application are typically stored inside its DEX files. Generally, these identifiers are meaningful and can be used to derive the internal implementation details of the application. 
The exposure of these identifiers thus imposes a security risk for many software vendors. Hence, the most prominent and widely used obfuscation technique in Android is identifier renaming.

\begin{figure}[]
  \begin{center}
    \includegraphics[width=0.5\textwidth]{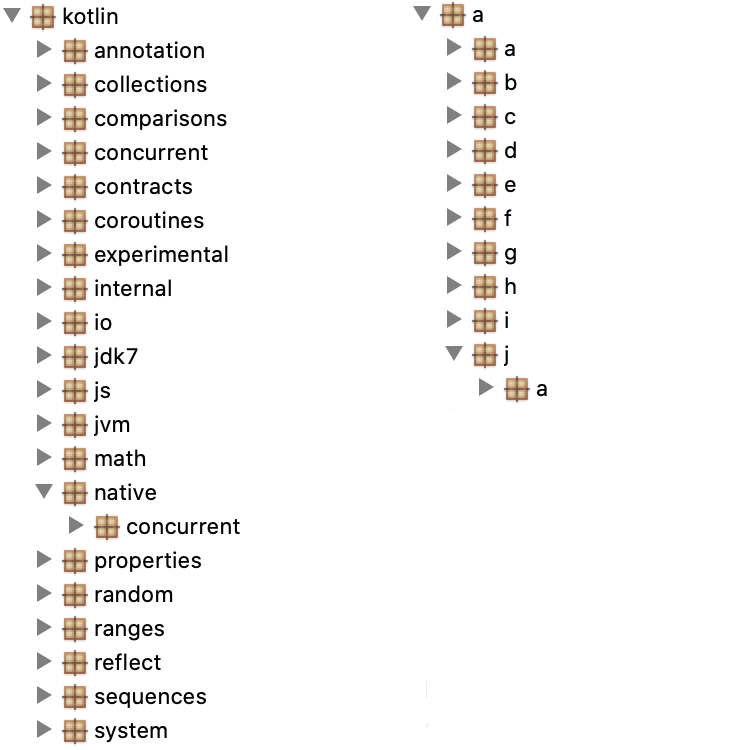}
  \end{center}
  \caption{Normal (left) and obfuscated (right) package names.}
  \label{fig:obf_packages}
\end{figure}

Identifier renaming can be done in different stages in the build process. ProGuard \cite{proguard}, arguably the most popular obfuscation tool works at source-code level \cite{androidobf}. It maps the original names to obfuscated names based on the user's configuration. Other tools such as DashO \cite{dasho} and DexProtector \cite{dexprotector} work directly on a DEX file. 
The package path and the identifier itself are frequently changed to short meaningless characters such as 'a' or 'aabb'. To what specific characters identifiers are being transformed in, usually depends on the tool or the configuration used. In Figure~\ref{fig:obf_packages}, an example of obfuscated package names is presented using ProGuard. On the left side of the figure, the original package names are presented. On the right side, the obfuscated package names are presented. Note that some packages are completely erased in the obfuscated version. This is because ProGuard also removes unused methods and packages, frequently resulting in significantly smaller applications.

\begin{figure*}[]
  \begin{center}
    \includegraphics[width=0.7\textwidth]{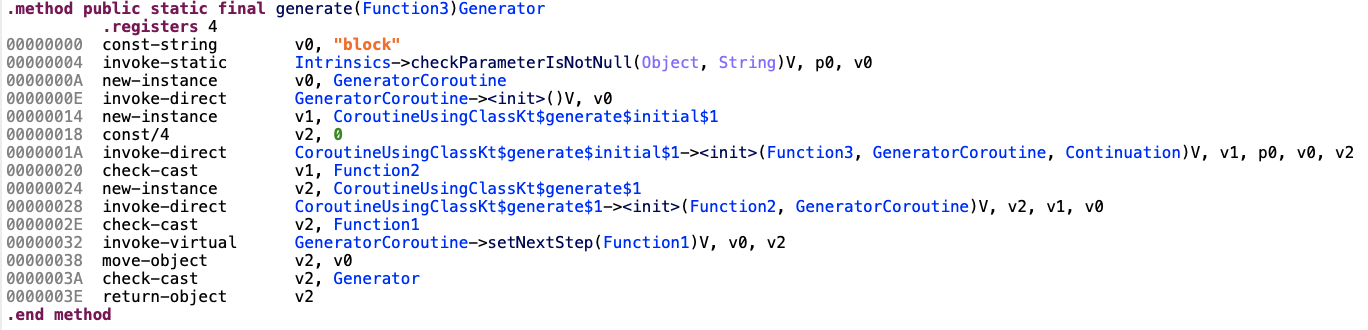}
  \end{center}
  \caption{Dalvik instructions of an example non-obfuscated method.}
  \label{fig:obf_method_org}
\end{figure*}

\begin{figure*}[]
  \begin{center}
    \includegraphics[width=0.4\textwidth]{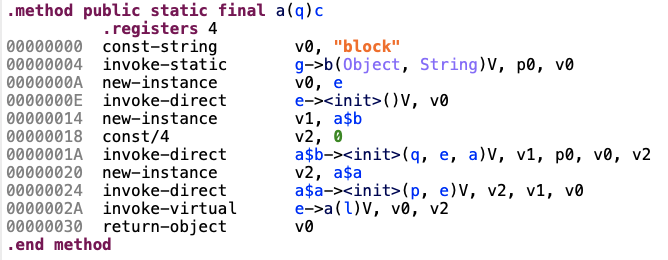}
  \end{center}
  \caption{Dalvik instructions of an example obfuscated method by ProGuard.}
    \vspace{-0.3cm}
  \label{fig:obf_method_obf}
\end{figure*}

In Figure \ref{fig:obf_method_org} and \ref{fig:obf_method_obf}, we present the Dalvik instructions of a method called \texttt{generate} and its obfuscated version respectively. Clearly, it is much harder to understand what the method does without knowing any of the original identifier names. The instructions in both figures are not completely equivalent because of ProGuard optimization.

Identifier renaming is considered a challenge for \emph{KotlinDetector} because it distorts the context of the application. This in turn results in complicating the task of extracting the features of the Kotlin language. The procedures used by \emph{KotlinDetector} to overcome the issues introduced by renamed identifier are discussed in Section~\ref{sec:implementation}.


\subsection{Compiled Kotlin in Android}
\label{sec:compiled_kotlin}
Any Android application developed in Kotlin uses the Android Runtime to handle execution. The resulting application after compiling is an APK archive containing DEX files. Just like Java, Kotlin is not directly compiled into native machine code, but compiles into the familiar Dalvik bytecode.

Most of the added functionality provided to a developer when using the Kotlin programming language, is implemented inside the Kotlin standard library. The Kotlin standard library is developed partially in Java and partially in Kotlin. The core of the standard library is developed mainly in Java. The standard library is fully open-source, just like the rest of Kotlin.

Many special operators in Kotlin are actually implemented by methods and fields defined in this library. Because of this, every application developed in Kotlin, must have at least some small part of the standard library in its bytecode. Hence, making it a good method of identifying usage of the language.

%% file: Related.tex
\section{Related Work}
\label{sec:related_work}

To the best of our knowledge, there has not been any extensive research on detecting Kotlin usage or extracting the language features from existing Android applications packages. Yet, some recent related studies do exist. In this section, we will review the relevant literature. 


Since Kotlin is a recently introduced language, several empirical studies have been conducted on its adoption by the developers. For instance, in \cite{Oliveira2020}, the authors employed a mixed method (mixing 9405 Stack Overflow questions with interviews of seven Android developers) to see what Kotlin language specific challenges are faced by the developers. Among their findings, Kotlin's \emph{null{-}safety} and inherent support for functional programming, as some of the most important novelties, are also the main sources of complications. The authors found out that the former may hinder interoperability with Java and the latter may compromise the code readability. Ardito et al.~\cite{Ardito20} conducted a similar study and obtained empirical evidence on code conciseness and lower number of \texttt{NullPointerExceptions} in Kotlin code. Another, perhaps more relevant study by Mateus and Martinez~\cite{studyqualitykotlin} performed an empirical study on the use of Kotlin programming language in open-source Android applications and their quality. Inside these open-source applications, they attempted to measure the amount of Kotlin code versus Java code with the conclusions: ``partial'', ``full'' or ``none''. They did so using two heuristics: (1) Checking whether there exists a directory named \texttt{kotlin} within the Android application's compressed archive. (2) Checking the amount of Kotlin present in the respective code repository of that application by using the GitHub API. For the latter, their tool inspects the most recent commit and counts the number of Kotlin bytes versus Java bytes in terms of source code size. It is unclear whether or not this statistic is equal to the statistic provided by GitHub Linguist, which we discussed earlier.

There are several problems with the approach of Mateus and Martinez. First of all, the first heuristic becomes useless if obfuscation tools such as ProGuard are used. This is due to the fact that the \texttt{kotlin} directory name is obfuscated and therefore cannot be easily identified. A weakness of the second heuristic is rather straightforward, it requires full access of the source code.
Unfortunately, many applications are still proprietary software and therefore their source code is not publicly available. Finally, one could argue GitHub Linguist already provides a near perfect solution to identify the programming language of a repository. \emph{KotlinExtractor} does not rely on the presence of any directory names which might become obfuscated. \emph{KotlinExtractor} also does not require full access to the application source code, it analyzes the Android package (APK).





Shah et al.~\cite{kotlinproguard} have studied the usage of Proguard, an obfuscator tool, to prevent App Repacking attacks. 
The authors showed that by using Proguard to obfuscate the published application, the decompiled code becomes much harder to read and understand for attackers. Hence, making it significantly more difficult to successfully perform an App Repacking attack.


%% file: Approach.tex
\section{Design}
\label{sec:approach}


The overarching goal of KotlinDetector is to detect the presence of Kotlin in Android applications and derive in-depth statistics about the usage of numerous language features. Towards achieving this goal, we define the following requirements:
\begin{enumerate}[label=R.\arabic*]
	\item \label{REQ1} Determine whether an application uses Kotlin at all.
    \item \label{REQ2} Determine whether an application employs obfuscation.
    \item \label{REQ3} (If Kotlin is used) How much of the Kotlin standard library is present.
    \item \label{REQ4} (If Kotlin is used) How much of the application consists of Kotlin code.
    \item \label{REQ5} (If Kotlin is used) Determine what language features are used.
    \item \label{REQ6} If a language feature is being used, determine how much it is used.
\end{enumerate}

In order to satisfy these requirements, we extract a set of features from a given APK file through invocation tracing, pattern scanning and parsing the manifest file. 
We now provide more details about these features and their mapping to the requirements.  

\subsection{Kotlin presence}
\label{subsec:kotlinpresence}
In order to satisfy~\ref{REQ1}, KotlinDetector checks if any of the standard library code of Kotlin is present in the compiled application (see Section~\ref{sec:compiled_kotlin}). Hence, identifying the presence of just one method, class or string reference from the Kotlin standard library inside the DEX file/s is already considered sufficient to label the application as having Kotlin presence. 

\subsection{Obfuscation detection}
The current version of \emph{KotlinDetector} supports mainly obfuscation based on identifier renaming since, according to~\cite{androidobf}, 57\% of Google Play Store applications employ this technique. In order to detect the presence of obfuscation (\ref{REQ2}) in an application, \emph{KotlinDetector} looks for the presence of Kotlin standard library package name (i.e. \texttt{kotlin}) in any type descriptor. The application is considered obfuscated even if chunks of the Kotlin standard library bytecode are present. 
An important thing to note here is that the internal structure of packages (the package tree) does not change when identifier renaming obfuscation is used.

\subsection{Percentage of Kotlin standard library code}
Since Kotlin code can be called from Java code smoothly, the first point of reference for computing the Kotlin usage in an application is to look for calls to Kotlin standard library. Thus, for satisfying \ref{REQ3}, KotlinDetector generates two distinct ratios between the total number of methods and classes of the standard library, and the overall number of methods and classes employed in a given application respectively. Let $M$ and $C$ denote the methods and classes in the application. The ratio on the number of the methods is obtained from $|M_{stdlib}| / |M|$ where $|M_{stdlib}|$ is the number of methods from the Kotlin standard library. Similarly, we obtained the ratio on the number of classes is obtained from $|C_{stdlib}| / |C|$.

\subsection{Percentage of Kotlin code}
\label{subsec:kotpercentage}
Now that we determined the presence of Kotlin, the use of obfuscation and the amount of Kotlin standard library calls in the application, we can determine how much of the code has been actually developed by using Kotlin (\ref{REQ4}). A naive approach to this is to determine the ratio between the number of Kotlin standard library invocations and the total number of invocations, i.e. $|I_{kotlin}|$ / $|I|$, through invocation tracing. One could even take a further step and calculate the ratio between the number of Kotlin classes and the total number of classes, formally $|C_{kotlin}| / |C|$. Here the assumption is each class with an invocation $i \in I_{kotlin}$ is marked as a Kotlin class.

However, this approach does not take the size of each class into consideration and some classes are significantly larger than the others. In order to resolve this issue, we computed the sum of the bytecode lengths of each method in a class. If $B(m)$ denotes the bytecode length of method $m$, we determined the number of Kotlin invocating bytes (\texttt{kot\_bytes}) as follows:
\[
\mathrm{kot\_bytes} = \sum_{c\in C_{kotlin}} \sum_{m\in M_c} B(m)
\]
By using a similar formula, we calculated the total number of bytes (denoted as \texttt{total\_bytes}) in the application. The ratio between these two measurements, we obtained a more accurate figure of Kotlin usage: 
\[
\mathrm{kot\_proj\_bytes\_ratio} = kot\_bytes ~/~ total\_bytes
\]

There was one problem with this ratio, however, that we were only interested in the code of the application itself rather than the dependencies and libraries used in the DEX files. 
Each Android application has structured package names. To gather an application's project packages, we need to filter the packages that start with the developers reversed domain name (\texttt{com.example}). This reversed domain name can be derived from the manifest file present in the APK file.

The second feature of \ref{REQ4} is concerning the ratio of Kotlin code in project classes. By identifying the project package suffix using the previous step, the ratio would then be equal to the size of Kotlin code in these classes divided by the size of these classes.




\subsection{Language features}
\label{sec:language_features}


We now turn our attention to \ref{REQ5} and \ref{REQ6}. Kotlin is a programming language that comes with a great set of language features, all specifically designed to enhance the development experience and the quality of the resulting code. Kotlin Web site~\cite{KotlinJava} lists 17 additional properties that Kotlin has compared to Java. Only some of these features are supported by KotlinDetector. In extraction of the language features, we exploit the fact that the Kotlin standard library is structured in such a way that common language features are incorporated in a separate package. Table I presents the features that can be detected by our tool and the respective package name. This structure gives us the ability to search for method invocation references towards these packages which can then be marked as the usage of the language feature in that package. For instance, for the feature \textit{ranges}, all the invoked references to the methods from the \texttt{kotlin.range} are retrieved. 
This works for most of the features except the \textit{Null Safety}. Features related to Null Safety such as safe calls, the let operator, the Elvis operator and safe casts are all visible in the bytecode of the compiled application, and KotlinDetector obtains the relevant information about the Null Safety from the bytecode.

\begin{table}[]
\centering
\begin{tabular}{| l | l |}
\hline
\textbf{Feature} & \textbf{Package} \\ 
\hline
Coroutines & \texttt{kotlin.coroutines} \\
\hline
Reflection & \texttt{kotlin.reflect} \\
\hline
Delegated Properties & \texttt{kotlin.properties} \\
\hline
Ranges & \texttt{kotlin.ranges} \\
\hline
Text & \texttt{kotlin.text} \\
\hline
Collections &  \texttt{kotlin.collections} \\
\hline
Comparisons & \texttt{kotlin.comparisons} \\
\hline
Concurrent & \texttt{kotlin.concurrency} \\
\hline
I/O & \texttt{kotlin.io} \\
\hline
Sequences & \texttt{kotlin.sequences} \\
\hline
\end{tabular}
\caption{\label{tab:features} Supported Kotlin Language Features and the Respective Packages}
\end{table}

Finally, even though popular obfuscation tools obfuscate these package names, we are able to determine which package the obfuscated package original name corresponds. We do this by using pattern scanning on a core method guaranteed to be present in the target package.

%% file: Implementation.tex
\section{Implementation}
\label{sec:implementation}
In this section the implementation details of \emph{KotlinDetector} and the technological choices are discussed. 

\subsection{Technology stack}
\label{sec:tech_stack}
To realize the full implementation of \emph{KotlinDetector}, several libraries and tools are utilized and incorporated. The application is developed from scratch in C++ including an efficient DEX-file parser. The C++ was the favorite choice of implementation since it guarantees fast performance and allows for jumping to specific offsets inside a file easily. The \emph{libzip} library created by D. Baron and T. Klausner \cite{libzip} is used to handle the APK files, which are essentially ZIP archives. To parse the \texttt{AndroidManifest.xml} file, which has a special XML encoding, the Android Binary XML Decoder library~\cite{axmlparser} is used. 

\begin{figure*}[!t]
  \begin{center}
    \includegraphics[width=\textwidth]{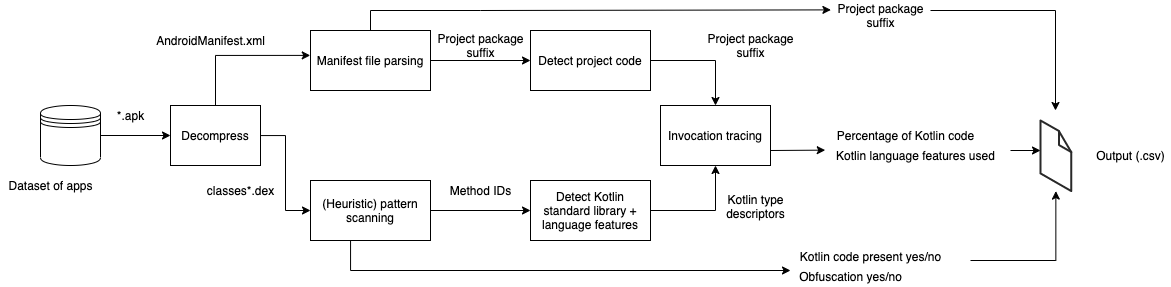}
  \end{center}
  \caption{The internal building blocks of KotlinDetector.}
  \label{fig:diagram}
\end{figure*}
\subsection{Overview}
Figure~\ref{fig:diagram} shows the components and the general flow of \emph{KotlinDetector} work to produce the desired results, in the following sections, more details are provided about these components.

\subsection{Scanning through Multiple DEX files}
To find all the DEX files inside an APK file, we search for file names matching the following pattern: \texttt{classes*.dex}.

\subsubsection{APK decompression}
In this step, the contents of an APK are decompressed and saved into memory. Each valid \texttt{classesN.dex} is located, decompressed, and saved to a byte array. Similarly, the content of the valid \texttt{AndroidManifest.xml} is also decompressed and saved into a byte array.
\subsection{AXML parsing}
The AXML parser walks through the byte array chunk by chunk computing a so-called Boost property tree to gather the application package. Extracting the application package still works even if the app is obfuscated using identifier renaming. It is because the paths that are included in the Android manifest file needs to match the ones found in the DEX file; otherwise, the operating system will not be able to locate the starting activity of an application.

\subsubsection{Pattern scanning}
The goal of this step is to search for the presence of a specific sequence of bytes inside the large DEX byte arrays, which is done using byte pattern scanning. This specific sequence of bytes is usually defined as a byte signature. A byte signature might represent a specific number, string or sequence of instructions in hexadecimal format. An example byte signature for the string \quotes{Hello} in ASCII would be \texttt{48 65 6C 6C 6F}.

\subsubsection{Heuristic method searching}
Any code that is part of the Kotlin standard library, shall start with \texttt{kotlin.*} inside its package name. Hence, a method, a field or a class can be verified as being part of the Kotlin standard library by checking if its type descriptor starts with the Kotlin standard library package path.
However, when an application is obfuscated by identifier renaming (i.e. when using ProGuard) the Kotlin standard library package path can be modified, which defies the above approach. Nevertheless, if we are able to determine the obfuscated package path we can still use this package path to check the type descriptor. This is possible due to the fact that the original package structure remains intact after applying the identifier renaming obfuscation.
So, a heuristic approach is employed here, which entails finding a method known to be part of the core of the Kotlin standard library to derive its package path.

\subsubsection{Invocation tracing}
After identifying the type of descriptor, a method's class needs to start with, in order to be considered as part of the Kotlin standard library or belong to a specific Kotlin language feature, the next step would be to trace the invoke instructions that references this method.
The idea of invocation tracing necessitates stepping through each instruction of each method of each class inside the application. If an instruction is an invoke instruction, the destination method is checked. Thus, all the methods of the application that being invoked would get traced.

\subsubsection{Feature extraction}
By iterating over all classes of an application and all methods of these classes, a class is marked as invoking Kotlin or invoking a Kotlin language feature, if one of its methods does.
Finally, the last step is determining all the ratio's defined in the previous section. Determining these ratios are just simple divisions based on the derived features using invocation tracing. 
The source code and the labeled dataset is available here \cite{allrec}

%% file: Evaluation.tex
\section{Evaluation}
\label{sec:evaluation}
Evaluating the \emph{KotlinDetector} tool is done at two levels: first, its ability to detect the presence of any Kotlin code in a given APK, second, its ability to extract the Kotlin language features. In doing so, numerous datasets have been purposely collected or built for this work, in addition, we used datasets from our previous studies. Evaluating the first capability is relatively easier than the second due to the existence of the GitHub Linguist tool and the public statements of many companies on their use of Kotlin in their mobile applications. On the other hand, evaluating the features extraction part is more challenging because there is no existing tool to compare against nor is there a labeled dataset. The GitHub Linguist tool does in fact provide the programming languages percentages for the files in a repository even though an exact match between these two tools is not expected. Nevertheless, it still gives a decent measure for an evaluation and thus a comparison is meaningful. 

\subsection{Comparing with GitHub Linguist}
\label{subsec:comparewithgithub}
As part of the evaluation study, the Kotlin percentages extracted by \emph{KotlinDetector} from the APK files are compared with those extracted by the GitHub Linguist tool from respective repositories. While the \emph{KotlinDetector} percentages reflect the ratio of Kotlin code versus Java code, the GitHub Linguist percentages take more languages (other than Java) into consideration. Thus, the output of both tools must be made comparable to each other. In order to achieve this, we use the following ratio: 

$$\mathrm{ling\_kot\_java\_ratio} = \frac{\mathrm{ling\_kot\_ratio}}{\mathrm{ling\_kot\_ratio}+\mathrm{ling\_java\_ratio}}$$


Note that $\texttt{ling\_kot\_ratio}$ and \texttt{ling\_java\_ratio} are taken directly from the GitHub Linguist output. In order to estimate the accuracy of \texttt{KotlinDetector}, we took the the absolute difference with \texttt{kot\_proj\_bytes\_ratio} (as given in Section~\ref{subsec:kotpercentage}): 

$$\mathrm{error}=|\mathrm{kot\_proj\_bytes\_ratio}-\mathrm{ling\_kot\_java\_ratio}|$$

If an application uses neither Java nor Kotlin, the \emph{ling\_kot\_java\_ratio} is set to zero. Other KotlinDetector's features are also being validated using this GitHub Linguist feature,  \texttt{ling\_kot\_java\_ratio}. For example, the Boolean feature that state whether the Kotlin standard library is present in an app or not. It was assumed to define the usage of the Kotlin programming language in an application. To validate this feature, the following statement is used:
\[
\mathrm{has\_kot\_github} =
\begin{cases}
    true & \text{if } \mathrm{ling\_kot\_java\_ratio}>0\\
    false              & \text{otherwise} \\
\end{cases}
\]

Here is the list of KotlinDetector's features that are validated using the GitHub Linguist tool:
\begin{itemize}
\item \texttt{kot\_proj\_bytes\_ratio}
\item \texttt{kot\_bytes\_ratio}
\item \texttt{kot\_proj\_classes\_ratio}
\item \texttt{kot\_classes\_ratio}
\item \texttt{kot\_proj\_invocations\_ratio}
\item \texttt{kot\_invocations\_ratio}
\end{itemize}
The results of the validation study show that \texttt{kot\_proj\_bytes\_ratio} has a relatively low error in comparison to other features. Thus, it is selected to evaluate the accuracy of \emph{KotlinDetector}.

\subsubsection{F-Droid Dataset}
\label{subsubsec:f-droiddataset}
For the purpose of comparing the results of \emph{KotlinDetector} with GitHub, we manually collected a sample of 100 Android applications from F-droid. We further collected the GitHub Linguist statistics of the latest commit of these apps in combination with the latest released APK file on GitHub. We 
uniformly randomly sampled these applications from the different categories found on F-droid. At the time of this research, F-Droid had 17 different application categories. Each released APK file contained in the dataset was confirmed to match with the corresponding GitHub commit and statistics. Hence, the sample was suitable to validate the performance of \emph{KotlinDetector}.

\begin{figure}[!t]
  \begin{center}
    \includegraphics[width=0.45\textwidth]{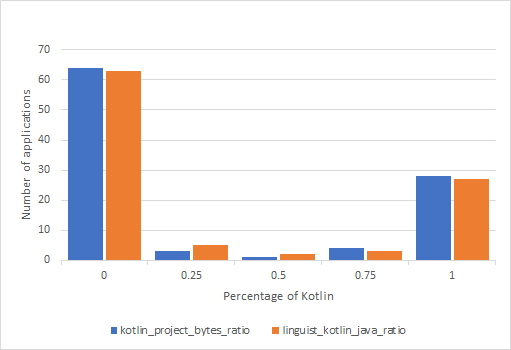}
  \end{center}
  \caption{The results of comparing KotlinDetector and Github Linguist with regard to the percentages of Kotlin in 100 applications.}
    \vspace{-0.5cm}
  \label{fig:hist_kotlin}
\end{figure}

In Figure~\ref{fig:hist_kotlin}, the summary of the results of scanning the 100 apps by both tools are presented. Overall, the two tools seem to be in a very good agreement in terms of the percentages of Kotlin in all apps. It is clear though that the two tools are in a strong agreement when an app does not have any Kotlin or has been totally written in Kotlin (100\%). A further look shows that there were no cases in which GitHub claimed the application had Kotlin traces, while \emph{KotlinDetector} claimed otherwise. Yet, there was few cases in which the opposite were true. Thus, it is safe to say that the output of the \emph{KotlinDetector} tool concerning Kotlin presence and percentages is considerably accurate.

\subsection{Additional Validation}
In this section, a number of datasets 
were used to validate the accuracy of the \emph{KotlinDetector} tool in detecting the presence of Kotlin.
\subsubsection{Initially handpicked Kotlin applications}
This dataset contains 201 handpicked Android applications that supposedly used Kotlin, based on vendors' public announcements. This dataset was initially used to develop the \emph{KotlinDetector}, giving direct access to a large number of varying Android applications developed in Kotlin. Out of the 201 applications, 70 applications found to use obfuscation by identifier renaming. Upon running \emph{KotlinDetector} on this dataset, 187 applications out of the 201 applications were found to use Kotlin. The 14 handpicked false positives and 14 randomly selected true positives applications were manually analyzed using JEB Android to validate the results of \emph{KotlinDetector} tool. The results that we obtained from this validation study assured the precision of our tool.

\subsubsection{Android applications from 2018 and 2020}
Two datasets of Android applications were collected at two different points in time. The first dataset, has 2,189 applications, was collected in 2018 \cite{2018dataset}, and the second dataset, has 1,970 applications, was collected in 2020.
Both datasets were uniformly randomly sampled from the Google Play Store.
Since Kotlin was introduced for Android in 2017 and its adoption has been growing even since, comparing the percentage of Kotlin applications in the 2018 dataset to the percentage of Kotlin applications inside the 2020 dataset should reflect that trend.
The results of running \emph{KotlinDetector} on both datasets showed that only 2.60\% of the applications in the data from 2018 used Kotlin in comparison to 22.39\% of the 2020 dataset. 
\subsubsection{Android malware applications 2013}
Lastly, \emph{KotlinDetector} was additionally tested on a malware dataset, that has 4,553 malware samples, collected in 2013 \cite{malwaredataset}. 
\emph{KotlinDetector} found no Kotlin traces in any of these applications. Any other result would not have been desirable, as, in 2013, Kotlin was not yet supported by the Android platform. This did not imply that no cases where \emph{KotlinDetector} would generate false positives existed. However, the finding provided additional validation for the tool, as the size of the dataset was large.

\subsection{Evaluating language feature detection}
\label{subsubsection:evalfeatures}
\begin{figure}[!t]
  \begin{center}
    \includegraphics[width=0.4\textwidth]{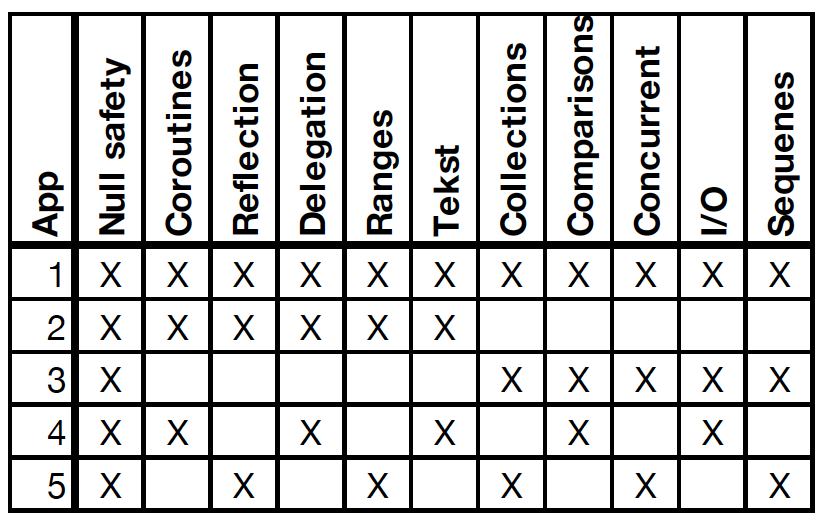}
  \end{center}
  \caption{Table containing the ground truth for the validation dataset.}
  \label{fig:table_features}
\end{figure}

Due to the lack of tools and labeled datasets that could be used to evaluate the accuracy of \emph{KotlinDetector} in extracting the set of language features, a selection of five different applications were developed. Each of these applications contained different Kotlin language features and varying implementations of these language features. The apps are further obfuscated using ProGuard using full identifier renaming. 
Figure~\ref{fig:table_features} describes the ground truth of the 10 applications and shows the list of Kotlin language features that are implemented in them.

\begin{figure}[!t]
  \begin{center}
    \includegraphics[width=0.4\textwidth]{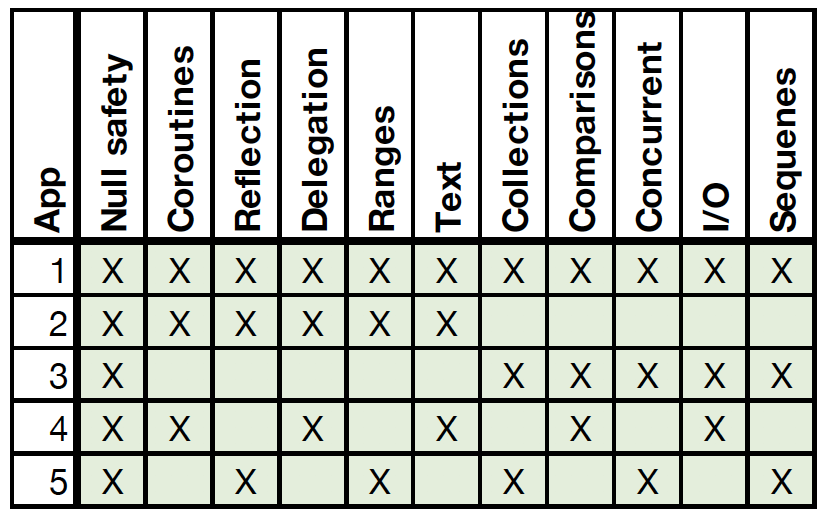}
  \end{center}
  \caption{Table presenting the presence of Kotlin language features detected by KotlinDetector for the non-obfuscated samples.}
    \vspace{-0.4cm}
  \label{fig:table_nonobf}
\end{figure}

In Figure~\ref{fig:table_nonobf} the output of \emph{KotlinDetector} for the non-obfuscated applications is presented. Correct identifications are marked in green. As can be seen, for all applications all the language features are correctly identified. This was expected, as all the original type descriptors are still available in each DEX file; therefore, each language feature could be detected with a high degree of certainty.

\begin{figure}[!t]
  \begin{center}
    \includegraphics[width=0.4\textwidth]{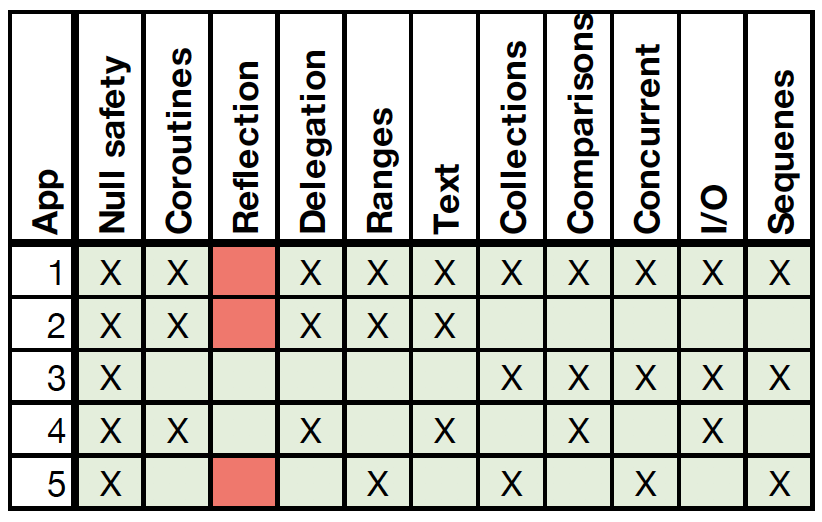}
  \end{center}
  \caption{Table presenting the presence of Kotlin language features detected by KotlinDetector for the obfuscated samples.}
  \label{fig:table_obf}
\end{figure}

In Figure~\ref{fig:table_obf}, the output of \emph{KotlinDetector} for the obfuscated applications is presented. Note that any incorrect or missing determination is marked in red. The only Kotlin language feature the tool missed identifying correctly was the use of the Kotlin reflection API. The Kotlin reflection API package (\texttt{kotlin.reflect}) due to the use of ProGuard, is found to contain only several interfaces. Since interfaces do not contain any code, a correct signature scanning to identify them was not possible. 

Apart from the use of the Kotlin reflection API, all other language features were correctly identified even when obfuscation by identifier renaming is used and the optimization of unused and dead code is enabled. Hence, it is safe to claim that the method of heuristically searching the obfuscated package path of a specific language feature to detect its use is a working approach of \emph{KotlinDetector}.

\subsection{Performance}
\label{subsection:performance}
In this section, the results of testing the performance of \emph{KotlinDetector} tool are discussed. The tool was used to scan the applications in the datasets that were discussed previously. The datasets, year of collection, percentage of apps that are using Kotlin, and average execution time per an app is displayed in Table~\ref{table:performance}. The data suggests that the apps that does not have Kotlin at all require minimum time because much of the search and tracing routines are being skipped. Another factor to be investigated here is the effect of the size of the application on the processing overhead.   
\begin{table}[]
\begin{tabular}{|c|c|c|c|c|}
\hline
\multicolumn{1}{|l|}{\textbf{Dataset}} & \multicolumn{1}{l|}{\textbf{Year}} & \multicolumn{1}{l|}{\textbf{Size}} & \multicolumn{1}{l|}{\textbf{\% Kotlin?}} & \multicolumn{1}{l|}{\textbf{Per App (seconds)}} \\ \hline
1                                      & 2013                               & 4553                               & 0                                        & 0.04                                            \\ \hline
2                                      & 2018                               & 2189                               & 2.6                                      & 0.32                                            \\ \hline
3                                      & 2020                               & 1970                               & 22.39                                    & 0.42                                            \\ \hline
4                                      & 2020                               & 100                                & 37                                       & 0.38                                            \\ \hline
5                                      & 2019                               & 201                                & 93                                       & 0.47                                            \\ \hline
\end{tabular}
\caption{\label{table:performance} The average processing time per an app for all datasets after running KotlinDetector.}
\end{table}

\begin{figure}[!t]
  \begin{center}
    \includegraphics[width=0.5\textwidth]{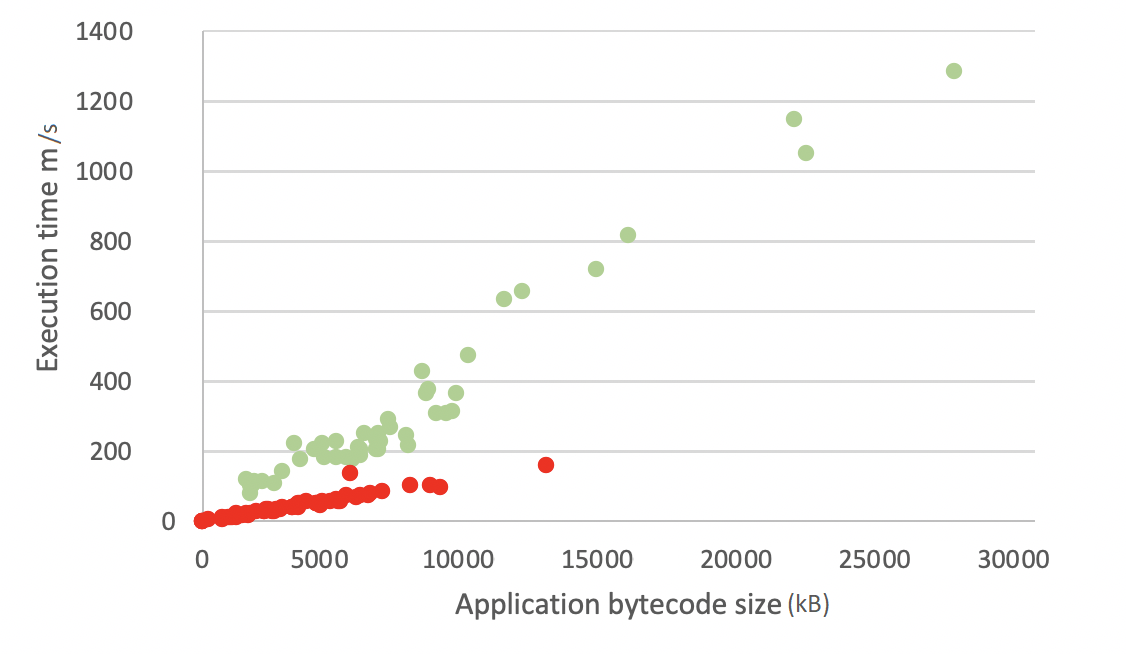}
  \end{center}
  \caption{Execution time plotted against application size. Red points indicate non-Kotlin applications, and green points indicate Kotlin applications.}
    \vspace{-0.4cm}
  \label{fig:plot_speed_split}
\end{figure}

In Figure~\ref{fig:plot_speed_split}, the execution time required to process an application is plotted against its size in bytecode using the F-Droid dataset \ref{subsubsec:f-droiddataset}. From the figure, the execution time required for analyzing an application depended on the size of the application in particular those containing Kotlin. 

A regression analysis was done which supports our analysis that the execution time depends on both the application size and whether or not the application uses Kotlin. Both variables had a significance value less than $0.05$. Further analysis was done to measure the effect of obfuscation on the execution time. The results showed that
the variable used to indicate whether or not the application is obfuscated had a significance value of $0.43$. Hence, the impact of obfuscation on the execution time was not significant. We can thus conclude that the required execution time depends only on the size of the application and whether or not the application is using Kotlin.

%% file: UseCase.tex
\section{Use Case Scenario}
\label{sec:usagescenario}
In this section, we present a usage scenario for our tool in which its output is combined with the output of \emph{AndroBugs}, a vulnerability scanning tool for Android applications\cite{AndroBugs15}, for vulnerability analysis over a number of datasets. Then, we perform correlation analysis to identify interesting relations (if any) among particular attributes of these datasets and vulnerabilities. Our aim is not to conduct an extensive analysis in this scenario but to present a setting where our tool can be used.

\subsection{Relevance}
\emph{KotlinDetector} tool is designed to help researchers identify and study closely Kotlin-based applications. Without this tool, it would not be easy for researchers to identify Kotlin-apps based solely on their APK files. Among all the studies that would benefit from this tool is the security and privacy. The real security and privacy implications of introducing Kotlin into Android mobile applications had not been largely investigated. Thus, in this use case, we are aiming at demonstrating how such a study can be conducted with the help of \emph{KotlinDetector} and AndroBugs. 
\subsection{Tools and Datasets}
AndroBugs is used to find predefined security vulnerabilities (also known as vectors) in Android
applications such as:
\begin{itemize}
\item SSL Connection Checking
\item Base64 String Encryption
\item WebView RCE Vulnerability Checking
\item KeyStore Protection Checking
\item KeyStore Type Checking
\item Runtime Command Checking
\item Runtime Critical Command Checking
\end{itemize}
AndroBugs detects four different categories of security vulnerabilities: critical, warning, notice and information. The output files of \emph{AndroBugs} are parsed and aggregated to facilitate the correlation study. For each application, we calculate the number of vulnerabilities in each of the four categories. We further add a feature that shows whether an app is using Kotlin or not.
The datasets that we use in this analysis are balanced datasets that we sample from the datasets that we presented earlier in Table~\ref{table:performance}.
The datasets are shown in Table~\ref{tab:correlations}. The first dataset is sampled from \emph{dataset 2} of Table~\ref{table:performance}, and the other two are sampled from \emph{dataset 3}. 
 
\subsection{Correlation}
In this analysis, we use Pearson test to measure the strength of association between whether an app is using Kotlin or not, the \emph{has\_kotlin\_stdlib} feature, which we obtain from\emph{KotlinDetector}, and the number of vulnerabilities in each of the four categories, which we obtain from \emph{AndroBugs}. The value of the correlation coefficient, goes from -1 to +1, is an indicative measure of the strength of the relationship. A value of $\pm 1$
indicates a very strong association, the association gets weaker as the value moves toward the zero. The sign of the value + or - implies the direction of the association, + means positive, and - means negative. In Table~\ref{tab:correlations} we show Pearson results of the three datasets. The results indicate either a strong or a medium correlation between the use of Kotlin and the four types of vulnerabilities: critical, warning, notice and information.
\subsection{Comparison}
\begin{figure}[!t]
  \begin{center}
    \includegraphics[width=0.4\textwidth]{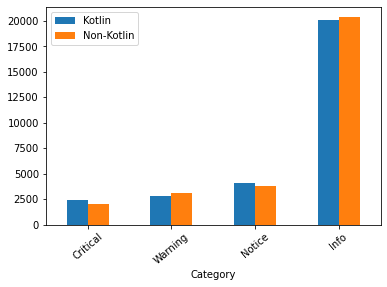}
  \end{center}
  \caption{A comparison between two datasets: 537 Kotlin and 537 non-Kotlin apps, in terms of the number of vulnerabilities in each of the four categories that AndroBugs detects.}
  \vspace{-0.4cm}
  \label{fig:table_obf}
\end{figure}

Furthermore, in Figure~\ref{fig:table_obf}, we show a comparison between two datasets, each contain 537 apps, in terms of the number of vulnerabilities in each of the four categories. Among the non-Kotlin apps, only 10 apps are found to contain no critical vulnerabilities and only 6 apps contain no warning vulnerabilities. On the other hand, among the Kotlin-based apps, only 39 apps are found to contain no critical vulnerabilities, only 36 apps contain no warning, and only a single app contains no warning. 

Finally, in Table~\ref{tab:vulnerabilities} we show the most frequent vulnerabilities found in both datasets. The critical and warning vulnerabilities are the same for both datasets; however, the notice and info vulnerabilities are different. The fact that 85\% of apps have a critical vulnerability such as not checking the SSL connection, says a lot about the state of mobile application insecurity. 

\subsection{Discussion}
The results above demonstrate the practical aspect of our tool in studying the security and privacy implications of using Kotlin in Android applications. Although, the datasets that we used were not large and only few features were considered, yet, the results are promising. We hope that our tool and labeled datasets will serve as a benchmark for further investigations in this field.

\begin{table}[]
\centering
\begin{tabular}{|l||*{4}{c|}}\hline
\backslashbox{\textbf{Size}}{\textbf{Feature}}
&\makebox[3em]{\textbf{\#Critical}}&\makebox[3em]{\textbf{\#Warning}}&\makebox[3em]{\textbf{\#Notice}}&\makebox[3em]{\textbf{\#Info}}\\\hline\hline
100 (50,50) &0.33&0.46&0.54&-0.51\\\hline
200 (100,100) &0.4&0.57&0.55&-0.55\\\hline
400 (200,200) &0.44&0.33&0.45&-0.46\\\hline
\end{tabular}
\caption{\label{tab:correlations} The correlations between the "has\_kotlin\_stdlib" feature of an app and the number of vulnerabilities in each category.}
\end{table}

\begin{table}[]
\setlength{\extrarowheight}{3pt}
\begin{tabular}{l|c|c|l|l|}
\cline{2-5}
\textbf{}                                                  & \multicolumn{1}{l|}{\textbf{No.}} & \multicolumn{1}{l|}{\textbf{Category}} & \textbf{Description}                            & \multicolumn{1}{c|}{\textbf{\%}} \\ \hline
\multicolumn{1}{|c|}{\multirow{4}{*}{\textbf{\rotatebox[origin=c]{90}{Kotlin}}}}     & 1                                      & Critical                               & SSL Connection Checking                         & 85\%                             \\ \cline{2-5} 
\multicolumn{1}{|c|}{}                                     & 5                                      & Warning                                & External Storage Accessing                      & 86\%                             \\ \cline{2-5} 
\multicolumn{1}{|c|}{}                                     & 14                                     & Notice                                 & File Unsafe Delete Checking                     & 98\%                             \\ \cline{2-5} 
\multicolumn{1}{|c|}{}                                     & 20                                     & Info                                   & SQLiteDB Transaction Deprecated  & 100\%                            \\ \hline
\multicolumn{1}{|l|}{\multirow{4}{*}{\textbf{\rotatebox[origin=c]{90}{Non-Kotlin}}}} & 1                                      & Critical                               & SSL Connection Checking                         & 85\%                             \\ \cline{2-5} 
\multicolumn{1}{|l|}{}                                     & 5                                      & Warning                                & External Storage Accessing                      & 95\%                             \\ \cline{2-5} 
\multicolumn{1}{|l|}{}                                     & 13                                     & Notice                                 & Android SQLite DBs Vulnerability & 98\%                             \\ \cline{2-5} 
\multicolumn{1}{|l|}{}                                     & 21                                     & Info                                   & Android SQLite DBs Encryption             & 100\%                            \\ \hline
\end{tabular}
\caption{\label{tab:vulnerabilities}The most frequent vulnerabilities found in the two app samples: 537 Kotlin-free apps and 537 Kotlin-based apps. }
\end{table}

%% file: Conclusion.tex
\section{Conclusion}
\label{sec:conclusion}
In an effort to understand the security and privacy implications of using Kotlin in Android applications, we developed the \emph{KotlinDetector} tool, available at \cite{allrec}, and the labeled dataset is available at \cite{datasetfl}. The tool is intended to detect the presence of Kotlin as well as extract its features from the APK files without requiring any access to the original source code. \emph{KotlinDetector} employs a number of techniques such as heuristic searching, pattern scanning, and invocation tracing to analyze the \emph{DEX} and \emph{AndroidManifest.xml} files inside an applications’ APK file. \emph{KotlinDetector} is capable of analyzing applications that use multi-DEX and are obfuscated using \emph{ProGuard}.

We evaluated the accuracy and performance of \emph{KotlinDetector} by running it over numerous datasets of applications and comparing it to the GitHub Linguist tool. Our results showed that analyzing an application using our tool is accurate and very efficient; can be completed within a matter of seconds. 
We also provided a usage scenario for our tool, in which its output is combined with the output of \emph{AndroBugs}, a vulnerability scanning tool, and a number of security and privacy implications of using Kotlin were identified.

Our evaluation shows promising results for the \emph{KotlinDetector} tool, which makes it a reliable source for conducting further studies related to the privacy and security of Kotlin-based mobile applications. We hope that our tool and labeled datasets will serve as a benchmark for further investigations in this field.

%% file: Ack.tex
\section{Acknowledgments}
\label{sec:ack}
We thank Dan Plamadeala (University of Groningen) and Job Heersink (University of Groningen) for the datasets that they collected or helped collecting.

%% file: paper.bbl
\begin{thebibliography}{10}
\providecommand{\url}[1]{#1}
\csname url@samestyle\endcsname
\providecommand{\newblock}{\relax}
\providecommand{\bibinfo}[2]{#2}
\providecommand{\BIBentrySTDinterwordspacing}{\spaceskip=0pt\relax}
\providecommand{\BIBentryALTinterwordstretchfactor}{4}
\providecommand{\BIBentryALTinterwordspacing}{\spaceskip=\fontdimen2\font plus
\BIBentryALTinterwordstretchfactor\fontdimen3\font minus
  \fontdimen4\font\relax}
\providecommand{\BIBforeignlanguage}[2]{{%
\expandafter\ifx\csname l@#1\endcsname\relax
\typeout{** WARNING: IEEEtran.bst: No hyphenation pattern has been}%
\typeout{** loaded for the language `#1'. Using the pattern for}%
\typeout{** the default language instead.}%
\else
\language=\csname l@#1\endcsname
\fi
#2}}
\providecommand{\BIBdecl}{\relax}
\BIBdecl

\bibitem{kotlinsupport}
\BIBentryALTinterwordspacing
T.~Verge. (2020, mar) Google is adding kotlin as an official programming
  language for android development. [Online]. Available:
  \url{https://www.theverge.com/2017/5/17/15654988/google-jet-brains-kotlin-programming-language-android-development-io-2017}
\BIBentrySTDinterwordspacing

\bibitem{kotlinpreferred}
\BIBentryALTinterwordspacing
TechCrunch. (2020, mar) Kotlin is now google’s preferred language for android
  app development. [Online]. Available:
  \url{https://techcrunch.com/2019/05/07/kotlin-is-now-googles-preferred-language-for-android-app-development}
\BIBentrySTDinterwordspacing

\bibitem{kotlinhome}
\BIBentryALTinterwordspacing
Google. (2020, mar) Kotlin and android. [Online]. Available:
  \url{https://developer.android.com/kotlin}
\BIBentrySTDinterwordspacing

\bibitem{fdroidstats}
\BIBentryALTinterwordspacing
IzzySoft. (2020, may) F-droid main repository. [Online]. Available:
  \url{https://apt.izzysoft.de/fdroid/?repo=main}
\BIBentrySTDinterwordspacing

\bibitem{F-DroidRepo}
\BIBentryALTinterwordspacing
IzzySoft and F-Droid. "f-droid main repository". [Online]. Available:
  \url{https://apt.izzysoft.de/fdroid/?repo=main}
\BIBentrySTDinterwordspacing

\bibitem{Playstorerepo}
\BIBentryALTinterwordspacing
statista. "number of apps available in leading app stores as of 3rd quarter
  2020". [Online]. Available:
  \url{https://www.statista.com/statistics/276623/number-of-apps-available-in-leading-app-stores/}
\BIBentrySTDinterwordspacing

\bibitem{linguist}
\BIBentryALTinterwordspacing
GitHub. (2020, jun) Github linguist - language savant. [Online]. Available:
  \url{https://github.com/github/linguist}
\BIBentrySTDinterwordspacing

\bibitem{studyqualitykotlin}
\BIBentryALTinterwordspacing
B.~G. Mateus and M.~Martinez, ``An empirical study on quality of android
  applications written in kotlin language,'' \emph{CoRR}, vol. abs/1808.00025,
  2018. [Online]. Available: \url{http://arxiv.org/abs/1808.00025}
\BIBentrySTDinterwordspacing

\bibitem{AndroBugs15}
\BIBentryALTinterwordspacing
Y.-C. Lin. (2015) Androbugs framework. [Online]. Available:
  \url{https://github.com/AndroBugs/AndroBugs{\_}Framework}
\BIBentrySTDinterwordspacing

\bibitem{obfdef}
\BIBentryALTinterwordspacing
S.~Schrittwieser, S.~Katzenbeisser, J.~Kinder, G.~Merzdovnik, and E.~Weippl,
  ``Protecting software through obfuscation: Can it keep pace with progress in
  code analysis?'' \emph{ACM Comput. Surv.}, vol.~49, no.~1, Apr. 2016.
  [Online]. Available: \url{https://doi.org/10.1145/2886012}
\BIBentrySTDinterwordspacing

\bibitem{androidobf}
\BIBentryALTinterwordspacing
S.~Dong, M.~Li, W.~Diao, X.~Liu, J.~Liu, Z.~Li, F.~Xu, K.~Chen, X.~Wang, and
  K.~Zhang, ``Understanding android obfuscation techniques: {A} large-scale
  investigation in the wild,'' \emph{CoRR}, vol. abs/1801.01633, 2018.
  [Online]. Available: \url{http://arxiv.org/abs/1801.01633}
\BIBentrySTDinterwordspacing

\bibitem{proguard}
\BIBentryALTinterwordspacing
Guardsquare. (2020, jul) Java obfuscator and android app optimizer. [Online].
  Available: \url{https://www.guardsquare.com/en/products/proguard}
\BIBentrySTDinterwordspacing

\bibitem{dasho}
\BIBentryALTinterwordspacing
PreEmptive. (2020, jul) Dasho java obfuscator \& android obfuscator. [Online].
  Available: \url{https://www.preemptive.com/products/dasho/overview}
\BIBentrySTDinterwordspacing

\bibitem{dexprotector}
\BIBentryALTinterwordspacing
DexProtector. (2020, jul) Mobile application security. [Online]. Available:
  \url{https://dexprotector.com/}
\BIBentrySTDinterwordspacing

\bibitem{Oliveira2020}
V.~Oliveira, L.~Teixeira, and F.~Ebert, ``On the adoption of kotlin on android
  development: {A} triangulation study,'' in \emph{{SANER}}.\hskip 1em plus
  0.5em minus 0.4em\relax {IEEE}, 2020, pp. 206--216.

\bibitem{Ardito20}
L.~Ardito, R.~Coppola, G.~Malnati, and M.~Torchiano, ``Effectiveness of kotlin
  vs. java in android app development tasks,'' \emph{Inf. Softw. Technol.},
  vol. 127, p. 106374, 2020.

\bibitem{kotlinproguard}
Y.~Shah, J.~Shah, and K.~Kansara, ``Code obfuscating a kotlin-based app with
  proguard,'' 02 2018, pp. 1--5.

\bibitem{KotlinJava}
\BIBentryALTinterwordspacing
K.~W. Site. (2021, feb) Comparison to java programming language. [Online].
  Available:
  \url{https://kotlinlang.org/docs/reference/comparison-to-java.html}
\BIBentrySTDinterwordspacing

\bibitem{libzip}
\BIBentryALTinterwordspacing
D.~Baron and T.~Klausner. (2020, jul) Coding conventions. [Online]. Available:
  \url{https://libzip.org/}
\BIBentrySTDinterwordspacing

\bibitem{axmlparser}
\BIBentryALTinterwordspacing
Y.~Tsutano. (2017, jul) Android binary xml decoder. [Online]. Available:
  \url{https://github.com/ytsutano/axmldec}
\BIBentrySTDinterwordspacing

\bibitem{allrec}
L.~Oosterhaven, ``{KotlinExtractor and Output File},''
  \url{"https://cutt.ly/TkKbpLk"}, February 2021.

\bibitem{2018dataset}
F.~Mohsen, H.~Abdelhaq, H.~Bisgin, A.~Jolly, and M.~Szczepanski, ``Countering
  intrusiveness using new security-centric ranking algorithm built on top of
  elasticsearch,'' 08 2018.

\bibitem{malwaredataset}
\BIBentryALTinterwordspacing
H.~Kang, J.~Jang, A.~Mohaisen, and H.~K. Kim, ``Detecting and classifying
  android malware using static analysis along with creator information,''
  \emph{CoRR}, vol. abs/1903.01618, 2019. [Online]. Available:
  \url{http://arxiv.org/abs/1903.01618}
\BIBentrySTDinterwordspacing

\bibitem{datasetfl}
F.~Mohsen and L.~Oosterhaven, ``{Labeled Dataset},''
  \url{"https://cutt.ly/6x2f5J7"}, February 2021.

\end{thebibliography}
